\documentclass[10pt, conference]{IEEEtran}

\usepackage{mathpazo}
\usepackage[scaled]{berasans}
\usepackage[scaled]{beramono}
\usepackage[euler-digits]{eulervm}
\normalfont
\usepackage{microtype}
\usepackage[utf8]{inputenc}
\usepackage[T1]{fontenc}
\usepackage[english]{babel}
\usepackage{amssymb,amsmath,amsthm}
\usepackage{mathtools}

\usepackage{listings}

\PassOptionsToPackage{hyphens}{url}
\usepackage{hyperref}

\usepackage[]{xcolor}

\title{The Matrix Chain Algorithm\\to Compile Linear Algebra Expressions}

\author{\IEEEauthorblockN{Henrik Barthels}
\IEEEauthorblockA{AICES, RWTH Aachen\\
barthels@aices.rwth-aachen.de}
\and
\IEEEauthorblockN{Paolo Bientinesi}
\IEEEauthorblockA{AICES, RWTH Aachen\\
pauldj@aices.rwth-aachen.de}
}

\begin{document}

\maketitle

\section{Introduction}

The need to translate linear algebra operations into efficient code arises in
a multitude of applications. For instance, expressions such as
$$b = S^H H^H \left(\sigma HH^H + Q \right)^{-1}r$$
and
$$x = \left( \Sigma^T \Sigma + D^2 \right)^{-1} \Sigma^T b$$
occur in information theory \cite{albataineh2014}, and regularization
\cite{noschese2016}, respectively. Given such expressions, 
we are interested in the automatic generation of code that is at least as fast
and as numerically stable as what an expert would produce.

Conceptually, the problem is similar to how compilers cast scalar
expressions in terms of the available instruction set.
The corresponding problem for linear algebra expressions (involving matrices)
is much more challenging, and requires expertise in both numerical
linear algebra and high-performance computing.
On the one hand, one wants to take advantage of highly optimized
building blocks for matrix  operations, such as those as provided by the
BLAS~\cite{dongarra1990} and LAPACK~\cite{anderson1999} libraries. 
On the other hand, transformations based on 
associativity, commutativity and distributivity play an essential role. 
Further complication comes from the fact that
matrices frequently have structures and properties that can be exploited both
to transform---and thus simplify---expressions, and to evaluate them more
efficiently. The application of this kind of knowledge affects not only the computational cost, but also the necessary amount of storage space, and numerical accuracy.

At the moment, there are two options for dealing with complex matrix
expressions. One either has to map the expressions to kernels manually, or
use high-level programming languages and environments such as
Matlab and R. The first option involves a lengthy, error-prone process that
usually requires a numerical linear algebra expert. The second option, using
high-level programming languages, is a very convenient alternative in terms of
productivity, but rarely leads to the same performance levels as code produced
by an expert. As a simple example, consider an expression containing the
inverse operator: in Matlab, this is directly mapped to an explicit matrix
inversion, even though a solution that relies on linear systems is usually
both faster and numerically
more stable; in
this case, it is up to the user to rewrite the inverse in
terms of the slash ({\tt/}) or backslash ({\tt\textbackslash}) operators, which
solve linear systems.
Products are another example: Let $M_1, M_2 \in \mathbb{R}^{n \times n}$, $x
\in \mathbb{R}^{n}$. Depending on whether $M_1 M_2 x$ is computed from the
left, that is, parenthesized as $(M_1 M_2) x$, or from the right ($M_1 (M_2
x)$), the calculation requires either $\mathcal{O}(n^3)$ or $\mathcal{O}(n^2)$ scalar operations. In Matlab, products are always evaluated from left to right~\cite{matlabdoc:short}.
In other high-level languages such as 
Mathematica~\cite{mathematicadoc:short} and Julia~\cite{bezanson2012}, the situation is analogous.

  \begin{figure}[tb]
    \begin{minipage}{\columnwidth} 
      \begin{align}
        \text{chain} &\rightarrow \text{factor} \cdot \text{chain} \mid \text{factor} \label{eq:rule1}\\
        \text{factor} &\rightarrow \text{op} \mid \text{op}^T \mid \text{op}^{-1} \mid \text{op}^{-T} \label{eq:rule2}\\
        \text{op} &\rightarrow \text{symbol} \mid \text{symbol}_\text{indices} \label{eq:rule3}\\
        \text{indices} &\rightarrow \text{index} \; \text{indices} \mid \text{index} \label{eq:rule4}
      \end{align}
    \end{minipage}
    \caption{Grammar describing the expression we are concerned with.
    }
    \label{grammar}
  \end{figure}

Our end goal is a compiler that takes a mathematical description of a linear
algebra problem and finds an efficient mapping onto high-performance routines offered by libraries. 
In this document, we are
concerned with the mapping of expressions consisting of products, as described by the grammar in Figure \ref{grammar} (e.g., $X :=
A B^T C$ and  $x := A^{-1} B y$, where $A, B, C, X$ are matrices, and $x$ and
$y$ are vectors), onto a set $K$ of computational kernels (e.g.: \texttt{C:=A*B},
\texttt{C:= A$^\text{\texttt -1}$*B}, \texttt{B:= A$^\text{\texttt -1}$},
\dots). For a given performance metric, we are interested in the optimal
mapping. This problem can be seen as a generalization of the matrix chain
problem: Given a \emph{matrix chain}, a product $M_1 \cdots M_k$ of matrices
with different sizes, the question is how to parenthesize it so that the result can be computed with the minimal number of scalar operations. 
Our approach uses an extended version of the $\mathcal{O}(n^3)$ dynamic
programming matrix chain algorithm presented in \cite{cormen1990}. We
refer to the problem as the ``Generalized Matrix Chain Problem''
(GMCP) and call the presented algorithm ``GMC algorithm''.

\section{Generalizations}

We extend the original matrix chain algorithm in four ways:

\paragraph{Operations}
The GMC algorithm is able to deal with the transpose and inverse as additional
operators. The combination of those operators
with the multiplication leads to a rich set of different expressions, for
example $A B^T$, $A^{-1} B$, and $A^{-1} B^{-T}$. While mathematically all those expressions can be computed as a composition of explicit unary operations ($X:= A^{-1}$ and $X:= A^{T}$) and a plain multiplication ($X:=AB$), this is in many cases not advisable for performance and stability reasons. The selection of the best sequence of kernels is done by a search-based approach inspired by the linear algebra compiler
CLAK~\cite{fabregat-traver2013a}.
\paragraph{Properties}

Many linear algebra operations can be sped up by taking advantage of the
properties of the involved matrices. For example, the multiplication of two
lower triangular matrices requires $n^3/3$ scalar operations, as opposed to $2n^3$ operations for the multiplication of two full matrices \cite{higham2008}. Furthermore, properties propagate with the application of kernels. Take the product $A B^T$ as an example. If $A$ is lower triangular and $B$ is upper triangular, it is possible to infer that the entire product is lower triangular as well. The GMC algorithm symbolically infers the properties of intermediate operands and uses those properties to select the most suitable kernels.

\paragraph{Cost Function}

The original matrix chain algorithm minimizes the number of scalar operations
(FLOPs) necessary to compute the matrix chain. In the GMC algorithm, we allow the use of an arbitrary metric, which could be performance (FLOPS/sec), numerical accuracy, memory consumption, or a combination of multiple objectives.

\paragraph{Indices}

The grammar (Figure \ref{grammar}) allows matrices to be annotated with indices.
Consider the assignment $X_{ij}:= A_i B C d_j$ as an example. Instead of one single chain,
a two-dimensional grid of chains has to be computed. Clearly, some segments
are common to multiple chains; for performance reasons it is therefore
beneficial to reuse them. The GMC algorithm is able to find the optimal solution for indexed chains like this one.

\section{The Algorithm}

\begin{figure}[]
\lstset{
	language=Python,
	columns=[c]fullflexible,
	mathescape=true,
	tabsize=2,
	breaklines=true,
	lineskip=1pt,
}
\hrule
\begin{lstlisting}
for $l \in \{1, \ldots, n-1\}$:
	for $i \in \{0, \ldots , n - l - 1\}$:
		$j := i + l$
		for $k \in \{i, \ldots, j-1\}$:
			op1 $:=$ tmps[$i$][$k$]
			op2 $:=$ tmps[$k+1$][$j$]
			[sequence, seq_cost] $:=$ find_sequence(op1, op2)
			r $:=$ index_range(op1, op2)
			cost $:=$ costs[$i$][$k$] $+$ costs[$k+1$][$j$] $+$ seq_cost $*r$
			if cost < costs[$i$][$j$]:
				tmps[$i$][$j$] $:=$ create_tmp(op1, op2)
				tmps[$i$][$j$].properties $:=$
 		   		                  infer_properties(op1, op2)
				sequences[$i$][$j$] $:=$ sequence
				costs[$i$][$j$] $:=$ cost
				solution[$i$][$j$] $:=$ $k$
\end{lstlisting}
\hrule
\caption{The GMC algorithm.}
\label{pseudocode}
\end{figure}

Figure \ref{pseudocode} shows the full algorithm. Its complexity is
$$\mathcal{O}(n^3(k^3 + \gamma + p))\text{,}$$
where $n$ is the length of the matrix chain, $k$ is the number of kernels,
$\gamma$ is the number of indices occurring in the chain and $p$ is the number of properties that
are considered. We stress that the $k^3$ term is an upper bound that will not be reached in practice.

\section{Conclusion and Future work}

We consider the GMC algorithm to be an important step towards the development
of a compiler for linear algebra problems that finds optimized mappings to
kernels by applying domain specific knowledge. In the future, we will address
problems like common subexpression elimination and memory allocation.

\addcontentsline{toc}{section}{References}
\bibliographystyle{naturemag}
\bibliography{PhD_bibliography}

\end{document}